\documentstyle[aps,epsfig,preprint]{revtex}
\begin{document}
\title{Folding a 2-D powder diffraction image into a 1-D scan: a new procedure.}
\author{Antonio Cervellino$^{a,b}$, Cinzia Giannini$^{b}$, 
Antonietta Guagliardi$^{b}$ and Massimo Ladisa$^{b,\star}$}
\address{
$^{a}$ Paul Scherrer Institute CH-5232, Villigen PSI, Switzerland\\
$^{b}$ Istituto di Cristallografia (IC-CNR), Via G. Amendola 122/O, I-70126, Bari, Italy}
\date{\today}
\maketitle
\begin{abstract} 
A new procedure aiming at folding a powder diffraction 2-D into a 1-D scan 
is presented. 
The technique consists of three steps: tracking the beam centre by means of a 
Simulated Annealing (SA) of the diffraction rings along the same axis, 
detector tilt and rotation determination by a Hankel Lanczos Singular Value 
Decomposition (HLSVD) and intensity integration by an adaptive binning algorithm. 
The X-ray powder diffraction (XRPD) intensity profile of the standard NIST Si 640c 
sample is used to test the performances. Results show the robustness of the method and its 
capability of efficiently tagging the pixels in a 2-D readout system by matching 
the ideal geometry of the detector to the real beam-sample-detector frame. 
The whole technique turns out in a versatile and user-friendly tool for the 
$2\vartheta$ scanning of 2-D XRPD profiles.
\end{abstract}
\noindent 
$^\star$ Corresponding author,
E\_mail: massimo.ladisa@ic.cnr.it,
Phone: +39 0805929166,
Fax: +39 0805929170

\newpage

\section{Introduction}

X-ray powder diffraction (XRPD) technique is nowadays 
a well known tool to study crystalline properties, which provide 
important information for applications in many fields 
such as crystal structure solution, microstructure characterization, 
phase quantification and, recently, also in nanotechnology. In these fields, 
charge-coupled device (CCD) and imaging plate (IP) detectors have been 
widely used for collecting X-ray diffraction images, expecially because 
of their fast readout system 
(Hanley \emph{et al.} 2005; Hammersley \emph{et al.} 1997, Muchmore 1999, 
Phillips  \emph{et al.} 2002, Suzuki \emph{et al.} 1999). 
When the analysis of the 1-D powder 
pattern is rather sofisticated, as for the Rietveld techniques, a reliable 
pre-processing aimed at enhancing the quality of XRPD data is required. 
When working with 2-D detectors, like a CCD or an IP system, a crucial step 
is the folding of the 2-D powder diffraction collected image into a 1-D scan. 
In the present paper a new procedure aiming at folding a powder diffraction 
2-D into a 1-D scan is presented.

\section{The method}

The ideal X-ray powder diffraction geometry is cylindrical along the beam-sample 
axis. The diffraction cone with an opening angle 2$\vartheta$ and the apex on 
the sample reaches the imaging plate (a CCD detector for 
instance). Ideally the imaging plate would be orthogonal to the beam-sample axis 
and it would collect a series of circles 
with radii corresponding to different momentum transfer: the larger the radius 
the higher the momentum transfer (\emph{i.e.} the resolution). Unfortunately the imaging 
plate axis rarely fits the ideal geometry and the circles become ellipses with an 
eccentricity $\epsilon$ and a rotation $\phi_0$ parametrizing tilt and rotation of 
the detector frame. Furthermore the beam rarely reaches the geometrical centre of 
the plate and a global shift of the rings is needed to perform a correct 2$\vartheta$ 
scanning.
\par\noindent
The problem reduces to the nonlinear fit of the parameters appearing in the 
cartesian equation of an ellipse: 
${\displaystyle x^2/a^2 + y^2/b^2 = 1}$, where $a/b$ is the 
major/minor semiaxis. For a generic 
beam-sample-plate geometry the following equations hold: 
\begin{eqnarray}
\label{eq:ellipse_1}
x &=& x_c + \Delta x_c + a\ \cos (\phi -\phi_0)/\sqrt{1+\epsilon\ \sin^2 (\phi -\phi_0)}\; , \nonumber \\
y &=& y_c + \Delta y_c + a\ \sin (\phi -\phi_0)/\sqrt{1+\epsilon\ \sin^2 (\phi -\phi_0)}\; ,
\end{eqnarray}
being $\Delta x_c, \Delta y_c$ the discrepancies along $x,y$ axes between the beam 
centre and the imaging plate one $(x_c,y_c)$. $\epsilon$ is the ellipse eccentricity $a^2/b^2-1$. 
The cartesian equation for the ellipse finally reads:
\begin{equation}
\label{eq:ellipse_2}
\frac{ \rho^2}{a^2} \left[ 1+\epsilon S^2 \right]  - 1 =
2\rho \left[ \frac{\Delta x_c\, C}{a^2} + \frac{\Delta y_c\, S}{b^2} \right]
- \left[ \frac{\Delta x_c^2}{a^2} + \frac{\Delta y_c^2}{b^2} \right]  ,
\end{equation}
where $\rho = \sqrt{(x-x_c)^2 + (y-y_c)^2}$, $S=\sin(\phi-\phi_{0})$, 
$C=\cos(\phi-\phi_{0})$,  and $\phi  =\phi_{0}+\arctan\left( 
(y-y_c)/(x-x_c)\right)$.
\par\noindent
In Eq.~(\ref{eq:ellipse_2}) the l.h.s. is $\mathcal{O}(\epsilon)$ while the r.h.s. is 
$\displaystyle{ \mathcal O\left(\sqrt{\frac{\Delta x_c\ \Delta y_c}{a\ b}} \right) }$ 
and, since usually 
$\displaystyle{ \epsilon \ll \sqrt{\frac{\Delta x_c\ \Delta y_c}{a\ b}}}$,
Eq.~(\ref{eq:ellipse_2}) finally decouples. It turns out that the problem of finding the rings 
centroid can be solved independently from the $\epsilon/\phi_0$ determination which is indeed 
a {\it fine tuning} with respect to 
the former task. Needless to say that the attempt of finding the rings 
centroid cannot be accomplished with by using the eq. (\ref{eq:ellipse_2}) itself 
where both $\epsilon$ and $\phi_0$ are unknown at this stage.
\par\noindent The decoupling of the two problems, \emph{i.e.} centroid hunting and ellipses fitting, is 
the main advantage of our method in comparison to the ones nowadays available in the literature 
(see Hammersley \emph{et al.}, 1996) where the full set of parameters of an equation analogous to 
Eq.~(\ref{eq:ellipse_2}) is computed.

\subsection{Beam centroid hunting}

This task is carried out by using a Simulated Annealing (SA) 
algorithm (see Metropolis \emph{et al.} 1958, Pincus 1970, 
Kirkpatrick \emph{et al.} 1982, Kirkpatrick \emph{et al.} 1983, Kirkpatrick 1984) 
to search for a minimum of the XRPD rings centroid position 
on the imaging plate. 
As known, the algorithm 
employs a random search which not only accepts changes that decrease 
objective function $f$, but also some changes that increase it. 
Changes decreasing $f$ are accepted with a probability 
$\displaystyle{p=\exp(-\delta f/T)}$, where $\delta f$ is the decrease 
in $f$ and $T$ is a control parameter, which by analogy with the original 
application is known as the system {\it temperature} irrespective of the 
objective function involved. Here 
the objective function $f$ is the sharpness (defined \emph{e.g.} 
as the inverse normalized variance) of the radial distribution 
function of the highest intensity rings recorded on the imaging plate and 
the system {\it temperature} is the domain size around the guess beam 
centre during the random search. 
$f$ is computed with respect to the guess rings centre. 
In Fig.~1 the sharpness of the radial distribution function is shown 
before (top right) and after (bottom left) the SA. The increasing of 
$f$ sharpness is dramatic 
and it follows the correct positioning of the beam centre onto the 
imaging plate (see the two spots on the top left image of Fig.~1).

\subsection{Detector tilt and rotation determination}

Once the beam centre has been positioned, \emph{i.e.} $\Delta x_c = \Delta y_c = 0$, 
the eccentricity $\epsilon$ and rotation $\phi_0$ computation is straightforward; 
infact the eq. (\ref{eq:ellipse_2}) reads 
$\displaystyle{1/\rho^2 \simeq 1/a^2 \times}$ 
$\displaystyle{\left[ 1+\epsilon\ \sin^2(\phi - \phi_0) \right]}$.
The detector tilt and rotation are determined 
by using a subspace-based parameter estimation method called 
Hankel Lanczos Singular Value Decomposition (HLSVD) technique 
(Laudadio \emph{et al.}, 2002). 
The HLSVD 
method works as follows.
Let us model the $1/\rho^2$ by samples $1/\rho^2_n$ collected at angles 
$\phi_n$, $n = 0, \ldots N-1$ as the sum of $K$ exponentially damped 
complex sinusoids
\begin{equation}
1/\rho^2_n \simeq \sum_{k=1}^K a_k \exp (-d_k\ \phi_n) \ \cos[2 \pi f_k 
\phi_n + \varphi_k] \;,
\label{eq:model}
\end{equation}
$a_k$ is the amplitude, $\varphi_k$ the phase, 
$d_k$ the damping factor and $f_k$ the frequency of the $k^{th}$
sinusoid, $k = 1, \ldots, K$, with $K$ the number of damped
sinusoids. This multiparametric functions family has been 
successfully used (Laudadio \emph{et al.}, 2002) for modelling 
NMR signals, because these functions are intrinsically related to the physical 
description of a NMR signal. 
It has been shown, however, that 
this family has a great approximating power for 
many kinds of signals, including some of the natural functions used to 
describe crystal diffraction patterns, namely the Dirichlet kernel, 
related to the the Laue interference function 
(see Beylkin {\&} Monz{\'{o}}n 2005, Sec.~5.1). Practical tests 
(Ladisa \emph{et al.} 2005a; Ladisa \emph{et al.} 
2005b; Cervellino \emph{et al.} 2005a; Cervellino \emph{et al.} 
2005b) have also confirmed their effectiveness. 
To estimate the parameters involved in the 
approximation we proceed as follows.
The $N$ data points defined in (\ref{eq:model}) are arranged 
(see Beylkin {\&} Monz{\'{o}}n 2005, Sec.~3) into
a Hankel matrix $H \stackrel{def.}{=} H_{L \times M}^{m\ l} = 1/\rho^2_{m+l}$, 
$m=0,\ldots,M-1$, $l=0,\ldots,L-1$ , with $L+M = N+1$ 
($M \simeq L \simeq N/2$). 
The spectral decomposition of the Hankel matrix is performed by SVD. 
SVD gives this method its main advantage, that is 
flexibility, since the number of parameters is not fixed and 
it is determined only by the desidered agreement level 
between the model and the real peak profile shape. 
SVD of the Hankel matrix is the decomposition 
$ H_{L \times M} = U_{L \times L} \Sigma_{L \times M} V^H_{M \times 
M}$, where $\Sigma = {\rm diag}\{\lambda_1, \lambda_2, \ldots, 
\lambda_r\}$, $\lambda_1 \geq \lambda_2 \geq \ldots \lambda_r$, 
$r= \min(L, M)$, $U$ and $V$ are orthogonal matrices and the 
superscript $H$ denotes the Hermitian conjugate. 
A \emph{fast} SVD decomposition is achieved by 
the Lanczos 
bidiagonalization algorithm with partial 
reorthogonalization. 
This algorithm, based on FFT, computes the two matrix-vector
products which are performed at each step of the Lanczos procedure
in $\mathcal{O}((L+M) \log_2(L+M))$ rather than in $\mathcal{O}(LM).$
In order to obtain the {\it signal} subspace, the matrix $H$ is
truncated to a matrix $H_{K}$ of rank $K$
$H_K = U_K \Sigma_K V^H_K$, where $U_K$, $V_K$, and $\Sigma_K$ 
are defined by taking the first
$K$ columns of $U$ and $V$, and the $K \times K$ upper-left matrix
of $\Sigma$, respectively. As subsequent step, the least-squares
solution of the following over-determined  set of equations is
computed ${\displaystyle V^{(top)}_K E^H \simeq V^{(bottom)}_K}$, 
where ${\displaystyle V^{(bottom)}_K}$ and ${\displaystyle V^{(top)}_K}$ 
are derived from $V_K$ by
deleting its first and last row, respectively.
The $K$ eigenvalues ${\hat z}_k$ of matrix $E$ are used to
estimate the frequencies ${\hat f}_k$ and damping factors ${\hat
d}_k$ of the model damped sinusoids from the relationship
$\displaystyle{
{\hat z}_k = \exp \left [ \left ( - {\hat d}_k + \imath 2 \pi
{\hat f}_k \right ) \Delta \vartheta \right ], \label{eq:eigen}
}$
with $k = 1, \ldots, K$. Values so obtained are inserted into the
model equation (\ref{eq:model}) which yields the set of equations
\begin{equation}
1/\rho^2_n \simeq \sum_{k=1}^K a_k \exp (-\hat d_k\ \phi_n) \ 
\cos[2 \pi \hat f_k \phi_n + \varphi_k] \;,
\label{eq:model_eigen}
\end{equation}
with $n = 0, \ldots, N-1$. The least-squares solution of
(\ref{eq:model_eigen}) provides the amplitude ${\hat a}_k$ and
phase ${\hat \varphi}_k$ estimates of the model sinusoids which are used 
in the next step.
\par\noindent
The eq. (\ref{eq:model_eigen}) and the l.h.s. of the eq. (\ref{eq:ellipse_2}) 
formally match for $K$=3 and the parameters $\epsilon$ and $\phi_0$ finally 
read: $\displaystyle{\epsilon = 2\ a_1/(a_0 - a_1)}$ and 
$\displaystyle{\phi_0 = (\varphi_1 - \pi)/2}$. In passing we also quote the 
major semiaxis of the ellipse $\displaystyle{a=1/\sqrt{a_0 - a_1}}$. Results 
of HLSVD are summarized in Tab. 1; for the sample test the average estimated 
values for $\epsilon$ and $\phi_0$ are respectively $(0.93 \pm 0.10)\times 10^{-3}$ 
and $82^\circ \pm 5^\circ$.

\subsection{Tagging pixels and radial integration}

We shall focus on the pixel tagging problem.
We do not apply data-reduction procedures (detector response, diffraction 
corrections, Bragg peaks removal, etc.) at this stage since we are not 
interested in them. Nonetheless to say that their inclusion will not affect 
our conclusions (see Hammersley \emph{et al.} 1996 for instance).
\par\noindent
Tagging pixels onto the 2-D detector is a crucial step in the XRPD intensity 
profiling problem. Infact the final XRPD intensity profile follows 
the radial integration over the ellipses domains. Pixels are actually 
labeled by an adaptive binning algorithm. Adaptive binning is an analysis technique 
that involves subdividing a signal into samples that are more homogeneous than 
the whole signal (see Sanders \emph{et al.} 2001 and references therein for instance). 
This technique reveals information about the structure of the 
signal. Here we have adopted the following scheme. The XRPD 2-D image is divided 
into a set of annuli with common centre and eccentricity/phase computed in the 
previous steps. Then each annulus is tested to see whether it meets some 
criterion of homogeneity (\emph{e.g.}, whether all the pixels in the annulus are within a specific 
dynamic range); here we have chosen a conservative criterion: the same statistical 
representation within each annulus, \emph{i.e.} the same number of pixels per annulus. More 
complicated criteria can be devised to account for other features 
(detector response and diffraction correction for instance). If an 
annulus meets the criterion, it is not divided any further. If it does not meet the 
criterion, it is subdivided again and the test criterion is applied to the new annuli. 
This process is repeated iteratively until each annulus meets the criterion. The result 
might have annuli of several different sizes. Finally the $2\vartheta$ scanning is achieved 
by integrating the 2-D XRPD intensity profile over the annular domains tagged in the 
previous step. The resulting intensity profile corresponding to the bottom right image 
of Fig. 1 is shown in Fig. 2. 

\section{Results and conclusions}

We have considered a new approach to the $2\vartheta$ scanning in powder 
diffraction experiments 
using 2-D detectors. Our method relies on the decoupling of 
two main problems: the beam centre positioning and the calculation of 
eccentricity and rotation of the set of ellipses recorded. The two tasks 
are respectively accomplished with by a Simulated Annealing and a Hankel 
Lanczos Singular Value Decomposition, two well known and feasible 
techniques widely applied in several optimization problems. Tagging pixels 
task is achieved by an adaptive binning algorithm by which the whole image 
is partitioned in a set of annuli with the same statistics. Integration 
over the annular domains finally provides the $2\vartheta$ intensity 
profile. This approach has been succesfully applied on a set 
of XRPD CCD images of samples of the standard NIST Si 640c collected with a 
KappaCCD Nonius diffractometer.
\vskip 10pt
\par Acknowledgements \par

M.L. is indebted with A. Lamura, T. Laudadio and G. Nico for deep discussions 
on HLSVD method.

\begin{figure}
\caption{Top left. Original XRPD pattern: the beam centres before and after SA are spotted on the image. 
Top right. Radial distribution function before SA.
Bottom left. Radial distribution function after SA.
Bottom right. Original XRPD pattern: the beam centre after SA together with few ellipses tagged by the HLSVD 
procedure are enlightened.}
\begin{tabular}{cc} 
{\includegraphics[width=0.45\textwidth]{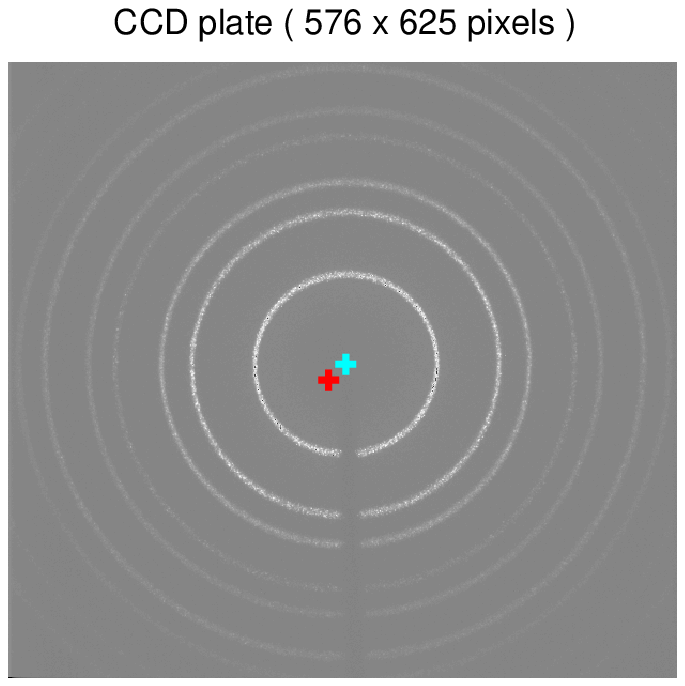}} 
& {\includegraphics[width=0.45\textwidth]{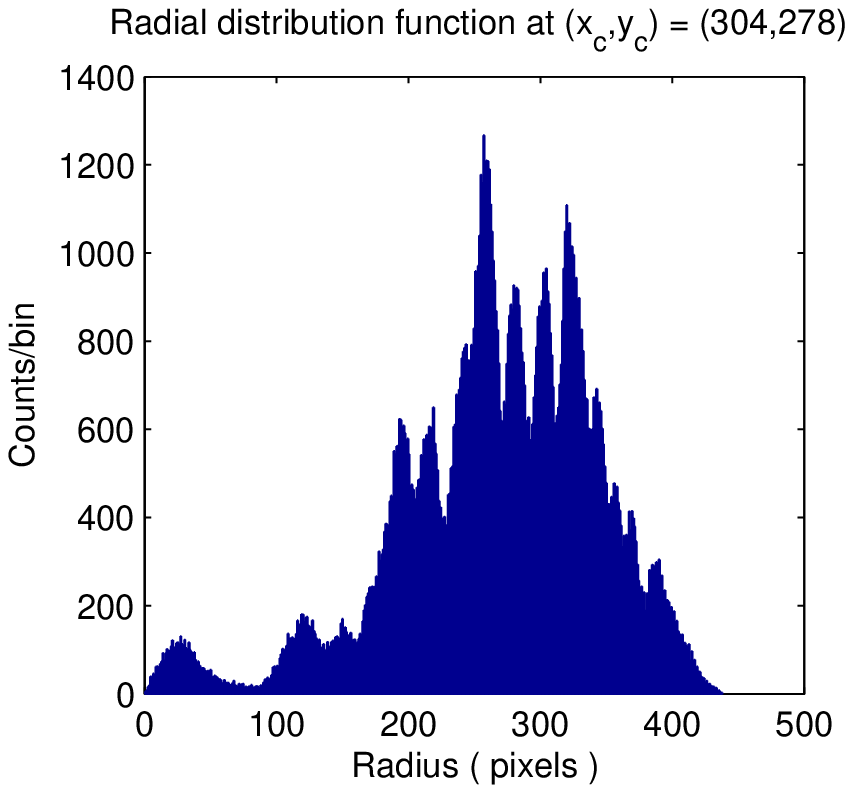}} \\
{\includegraphics[width=0.45\textwidth]{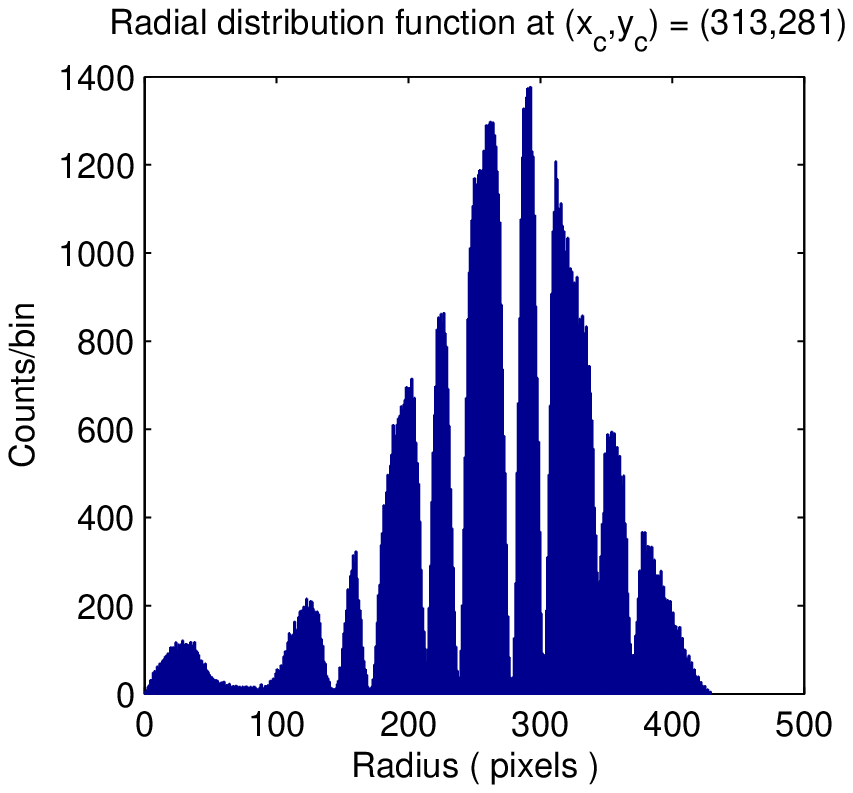}} 
& {\includegraphics[width=0.45\textwidth]{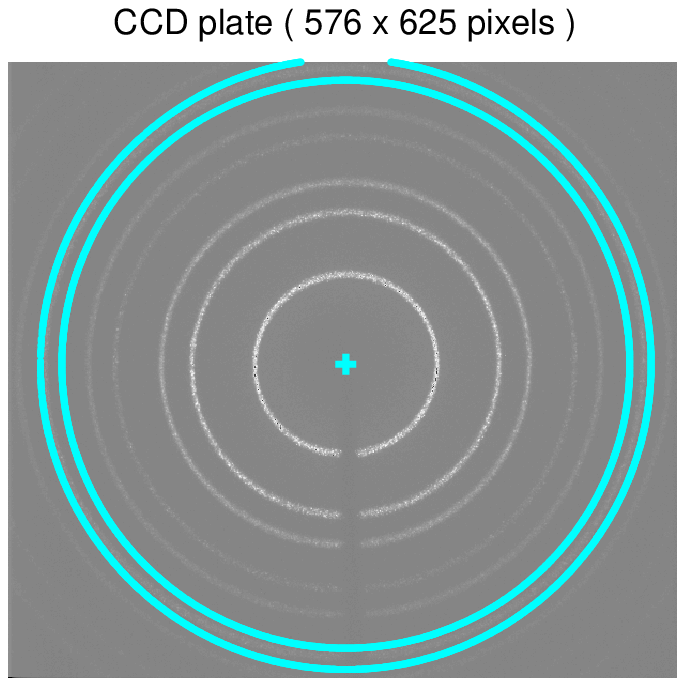}} \\
\end{tabular}
\end{figure}

\begin{table}
\caption{Parameters estimation by HLSVD ($K$=3) for few ellipses.}
\begin{tabular}{l|ccccccc} 
 parameters   & $a_0 \times 10^4$ & $a_1 \times 10^8$ & $\varphi_1$ & \vdots 
 & $\epsilon \times 10^2$ & $\phi_0$    & $a$   \\
\hline
 ellipse 1    & 0.117             & 0.268             & 319$^\circ$    & \vdots 
 & 0.091                  & 69.6$^\circ$   & 293 \\
 ellipse 2    & 0.119             & 0.247             & 342$^\circ$    & \vdots 
 & 0.083                  & 80.1$^\circ$   & 290 \\
 ellipse 3    & 0.120             & 0.291             & 346$^\circ$    & \vdots 
 & 0.097                  & 82.8$^\circ$   & 289 \\
 ellipse 4    & 0.146             & 0.395             & 357$^\circ$    & \vdots 
 & 0.108                  & 88.6$^\circ$   & 262 \\
 ellipse 5    & 0.145             & 0.325             & 346$^\circ$    & \vdots 
 & 0.089                  & 82.9$^\circ$   & 263 \\
\hline
\end{tabular}
\end{table}

\begin{figure}
\caption{XRPD intensity profile corresponding to the bottom right image of Fig. 1.}
\begin{center}
{\includegraphics*[width=1.05\textwidth]{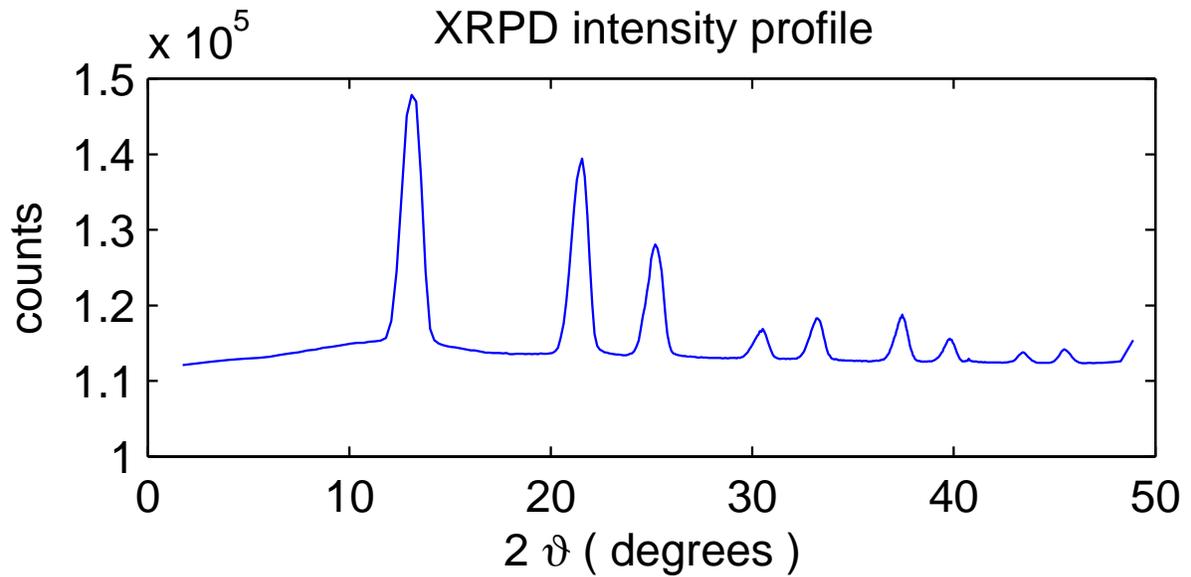}}
\end{center}
\end{figure}

\end{document}